\documentstyle[aps,prb,multicol,epsfig]{revtex}
\begin{document}
\title{
Collective excitations  in ferrimagnetic  Heisenberg  ladders}
\author{N. B. Ivanov~\cite{paddress} and J. Richter }
\address{
Institut f\"ur Theoretische Physik, Universit\"at Magdeburg,\\
P.O. Box 4120, D-39016 Magdeburg, Germany
}
\date{\today}
\maketitle
\begin{abstract}
We study ground-state  properties and the low-lying
excitations of Heisenberg  spin ladders  composed of 
two ferrimagnetic chains with alternating 
site spins  $(S_1>S_2)$  by using  the bosonic 
Dyson-Maleev formalism and
Lanczos numerical techniques. The emphasis is on
properties of the ferrimagnetic phase  which is stable 
for antiferromagnetic interchain couplings $J_{\perp}\geq 0$.
There are two basic implications of the underlying
lattice structure: (i) the spin-wave  excitations 
form folded acoustic and optical branches in 
the extended Brillouin zone  and
(ii) the  ground state parameters (such as the
on-site magnetizations  and 
spin-stiffness constant)  show a crossover behavior
in the weak-coupling region $0<J_{\perp}<1$.
The above peculiarities of the ladder ferrimagnetic state
are studied  up to second order in the quasiparticle
interaction and by a numerical diagonalization of 
ladders  containing  up to  $N=12$ rungs.
The presented results for the ground-state parameters
and the excitation  spectrum  can  be used in studies 
on  the low-temperature thermodynamics of ferrimagnetic 
ladders.  
 \end{abstract}
\draft
\pacs{PACS: 75.10.-b, 75.10.Jm, 75.40.Gb, 75.50.Gg}
\begin{multicols}{2}
\section{Introduction}
There is an impressive  difference between the properties
of a single spin-$\frac{1}{2}$ antiferromagnetic Heisenberg
chain and the related ladder system composed of two coupled
chains. The spin-$\frac{1}{2}$
chain is critical, with  a power-law decay of 
the spin correlations, whereas the spin-$\frac{1}{2}$ ladder has  
a finite energy gap in the excitation spectrum and exponentially 
decaying correlations.~\cite{dagotto} Similar behavior has
been found in multi-leg spin-$\frac{1}{2}$
ladders  with odd and even number of chains,
respectively.~\cite{white}
Since in spin  ladders the excitation spectrum is controlled by 
two energy scales (the  intrachain exchange constant $J$ and the 
transverse interchain exchange coupling $J_{\perp}$),
variation of  $J_{\perp}$ and the  applied   magnetic field
may produce a rich variety of specific quantum effects and
phase transitions. In particular, intermediate plateaus in
the magnetization processes
of a number of ladder systems have been predicted
and experimentally observed in the last few
years.~\cite{totsuka,giamarchi}

\begin{figure}
\centering\epsfig{file=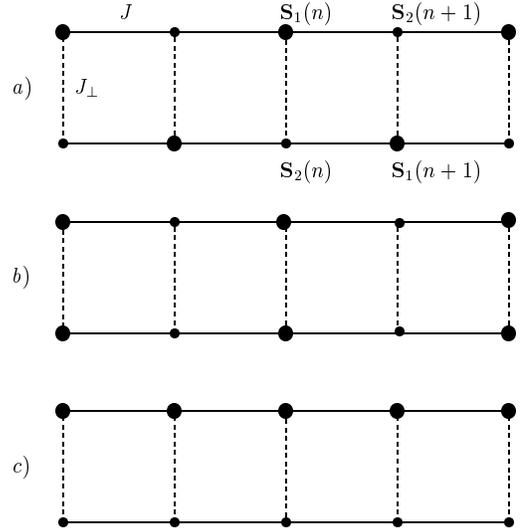,width=7cm}
\vspace{0.3cm}
\narrowtext
\caption{
Typical bipartite mixed-spin ladders composed of two types of
site spins  ${\bf S}_1(n)$ and ${\bf S}_2(n)$:
${\bf S}_1^2(n)=S_1(S_1+1)$, 
${\bf S}_2^2(n)=S_2(S_2+1)$ ($S_1>S_2$).
}
\label{ladders1}
\end{figure}

Synthesized quasi-one-dimensional
mixed-spin compounds \cite{kahn} constitute an
appropriate  base for future developments
in the physics of spin ladders. Most of them
are  molecular magnets containing two 
different transition-metal magnetic
ions alternatively distributed
on the lattice. Published experimental work
implies that the magnetic properties of these
mixed-spin materials are basically  described by
quantum Heisenberg spin models with antiferromagnetically
coupled nearest-neighbor localized spins.
Some typical examples of bipartite ladder structures
composed of two different spins are shown in
Fig.~\ref{ladders1}.
The first two ladder structures   reproduce, e.g.,
arrangements of the magnetic atoms   Mn
($S_1=\frac{5}{2}$) and Cu ($S_2=\frac{1}{2}$)
along the $a$-axis in the compounds
MnCu(pbaOH)(H$_2$O)$_3$ (pbaOH = 2-hydroxy-1,3-
propylenebisoxamato) and
MnCu(pba)(H$_2$O)$_3\cdot$2H$_2$O (pba =
1,3-propylenebisoxamato), respectively.
Along the $c$-axis the magnetic ions
in both mixed-spin compounds are arranged as shown in
Fig.~\ref{ladders1}(c).\cite{kahn}  Since the
molecular chemistry  admits a relatively
easy control of the molecular-unit positions,  it may be
expected that the discussed mixed-spin
ladders  will be synthesized in the near future.

Ground-state properties of the Heisenberg spin models
defined on the ladders shown in 
Fig.~\ref{ladders1}
have recently been studied  by using  semiclassical nonlinear 
$\sigma$ model 
techniques \cite{fukui,koga1} and strong-coupling expansions.
\cite{langari1,langari2} 
It was  found that an interplay between bond and
spin alternations can produce various
gapful  phases separated by critical lines
on the phase diagram.  
In this article we  address the 
mixed-spin ladder model  which is shown in Fig.~\ref{ladders1}(a) and  
defined  by the  Heisenberg Hamiltonian   
\begin{eqnarray}\label{h}
{\cal H}&=& J\sum_{n =1}^{N}
{\bf S}_1(n)\cdot {\bf S}_2(n+1)+{\bf S}_2(n)\cdot
{\bf S}_1(n+1)\nonumber\\
&+& J_{\perp}\sum_{n =1}^{N} {\bf S}_1(n)\cdot {\bf S}_2(n)\, .
\end{eqnarray}
The integers $n$ label  $N$ rungs, each of them  containing
two different site spins with quantum
spin numbers $S_1>S_2$. Below we frequently use the notations  
$\sigma = S_1/S_2>1$ and $S_2= S$, and set $J = 1$.            

For antiferromagnetic interchain couplings
($J_{\perp}>0$), the model (\ref{h}) exhibits 
a ferrimagnetic ground state characterized by the net
ferromagnetic moment $(S_1-S_2)N$.\cite{lieb} 
Such a magnetic phase  may be referred to as 
{\em a quantized unsaturated
ferromagnetic phase}: it is characterized both by
the quantized ferromagnetic order parameter
${\bf M}= \sum_{n=1}^N \left[{\bf S}_1(n)+
{\bf S}_2(n)\right]$
[quantized in integral or half-integral
multiples of the number of rungs, $M=(S_1-S_2)N$] and by
the macroscopic sublattice  magnetizations
$ M_A=\sum_{n=1}^N\langle S_1^z(n)\rangle$
and $M_B=\sum_{n=1}^N\langle S_2^z(n)\rangle$.
In the limit $J_{\perp} \rightarrow \infty$ the rung spins
form  local spin states with  a total spin $S_{tot}=S_1-S_2$: 
in particular, the $(1,\frac{1}{2})$ ladder  is equivalent to 
the  spin-$\frac{1}{2}$  ferromagnetic Heisenberg  
model with an  effective exchange  interaction 
$J_{eff}=-\frac{8}{9}$. On the other hand, 
for ferromagnetic interchain couplings  ($J_{\perp}<0$), 
it may be  generally expected  a magnetically 
disordered  spin-singlet ground state,   
since the isotropic Heisenberg model (\ref{h}) is
defined on a bipartite one-dimensional (1D) lattice.
          
In spite of the fact that the   ferrimagnetic
long-range order exists already in the limit of noninteracting
chains, in the antiferromagnetic 
region ($J_{\perp}>0$) the checkerboard  ladder  exhibits various 
specific  properties deserving a special study.
As shown below, the most interesting region  corresponds
to relatively small interchain couplings ($0<J_{\perp}<1$)  
where a number of ground-state parameters exhibit a 
crossover behavior: the latter is marked
by the extreme values of these parameters.  For instance, 
in that region the sublattice magnetization $M_A$ and 
the spin-stiffness constant $\rho_s$ reach their  maximal values. 
Another specific property of the ferrimagnetic ladder  
concerns the collective  spin-wave excitations,  
which form in the extended Brillouin zone $-\pi/a_0 <k <\pi/a_0$
folded acoustic  and optical branches characterized by  
a  minimum at the zone boundary
$k=\pi/a_0$ and a  maximum between $k=\pi /(2a_0)$ 
(for $J_{\perp}=0$)  and $k=\pi /a_0$ 
(for $J_{\perp}=\infty$),  $a_0$ being the lattice spacing
along the ladder. A purpose of the present work is
to  analyze the above mentioned peculiarities of the ladder system. 
                                      
The existence of a magnetic ground state opens the possibility 
of using systematic spin-wave  approaches  
for the 1D  Hamiltonian (\ref{h}). 
Indeed, recently published  
calculations\cite{ivanov2,yamamoto1,ivanov3} 
argued that the spin-wave theory (SWT) is capable to  
produce precise quantitative
results both for the ground-state parameters
and excitation spectrum of ferrimagnetic 
Heisenberg chains, including  the extreme quantum system 
$(1,\frac{1}{2})$.
Turning  to the  ferrimagnetic ladders,  
a SWT description  in the  weak-coupling limit 
$J_{\perp}\ll 1$  looks  to some extend puzzling since 
in that region  one may expect strong  
fluctuations of the macroscopic chain magnetizations around
the common  quantization axis.
Similar problems arise when the semiclassical 
nonlinear $\sigma$ model technique is applied to 
antiferromagnetic spin ladders.\cite{senechal}
A successful treatment of the weak-coupling limit $J_{\perp}\ll 1$
in ferrimagnetic ladders requires an appropriate choice
of the free-quasiparticle bosonic Hamiltonian. Using as a guide
the Lanczos numerical technique, it is argued below that
the Dyson-Maleev formalism is capable to
produce  precise quantitative description of the 
ferrimagnetic state in a relatively large  range of interchain
couplings, including the weak-coupling limit $J_{\perp}\ll 1$.

The paper is organized as follows. In Section 2 we introduce 
the bosonic representation of ${\cal H}$ and discuss
the choice of an appropriate free-quasiparticle
bosonic Hamiltonian. As a criterion we use the 
perturbation series for $M_A$ compared to Lanczos numerical results. 
In Section 3 we study the  excitation spectrum of the model 
up to second order in  the bosonic interaction $V$. 
Section 4 contains a summary of the results and
discussions concerning also the  disordered phase.  
In the Appendix we  calculate the second-order  
${\cal O}(V^2)$  corrections for  $M_A$.  

\section{Bosonic representation of ${\cal H}$: 
perturbation series for $M_A$}
\subsection{Bosonic Hamiltonian}
\begin{figure}
\centering\epsfig{file=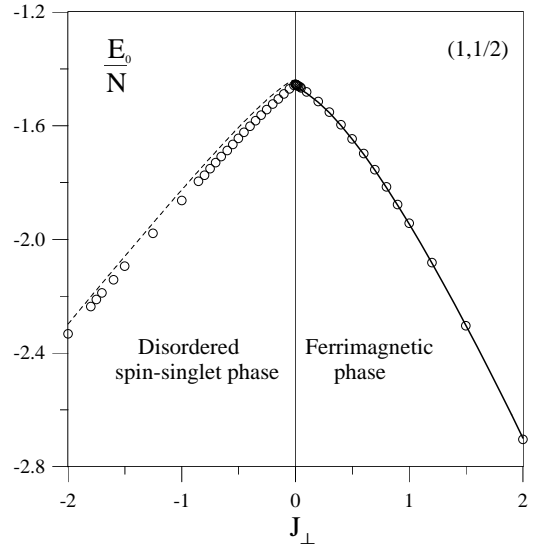,width=7cm}
\vspace{0.3cm}
\narrowtext
\caption{
Ground-state energy of the $(1,\frac{1}{2})$ ladder 
as a function of the interchain coupling $J_{\perp}$. 
The solid line displays the first-order 
${\cal O } (1/S)$  spin-wave result for the ferrimagnetic state, 
Eq.\ (\ref{e0}). The dashed line indicates the ground-state
energy of the disordered spin-singlet phase, as  obtained from 
the LSWT.
The  Lanczos numerical  results for a ladder  with $N=10$ rungs
are denoted by open circles. At $J_{\perp}=0$ there is
a well pronounced cusp in the data indicating a sharp
transition from the ferrimagnetic state into a  
disordered spin-singlet state.
}
\label{e01}
\end{figure}

We adopt the Dyson-Maleev formalism.~\cite{harris}
Performing subsequently the Dyson-Maleev, Fourier, and 
Bogoliubov transformations, 
we  obtain the following  Hamiltonian in terms of the 
quasiparticle bosonic operators 
$\alpha_k$ and $\beta_k$:\cite{note1}
\begin{equation}\label{hb}
 {\cal H}_B =E_0+{\cal H}_0+V\, ,
 \hspace{0.5cm}   V=V_2+V_{DM}\, .
\end{equation}
$E_0$ is the ground-state energy of the ferrimagnetic state
calculated up to first order in $1/S$ (see Fig.~\ref{e01}):
\begin{eqnarray}\label{e0}
\frac{E_0}{N}&=&-\left( 1+\frac{J_{\perp}}{2}\right)
\left[ 2\sigma S^2+S(1+\sigma)\left(
1-\frac{1}{N}\sum_k\varepsilon_k\right)\right]\nonumber \\
&-&2(c_1^2+c_2^2)-J_{\perp}(c_1^2+c_3^2)\nonumber \\
&-&(2c_2+J_{\perp}c_3)c_1\frac{\sigma +1}{\sqrt{\sigma}}\, ,
\end{eqnarray}
where
\begin{eqnarray}\label{c}
c_1&=&-\frac{1}{2}+\frac{1}{2N}\sum_k\frac{1}
{\varepsilon_k}\, ,\hspace{0.3cm}
c_2=-\frac{1}{2N}\sum_k\gamma_k\frac{\eta_k}
{\varepsilon_k}\, ,\nonumber \\
c_3&=&-\frac{1}{2N}\sum_k\frac{\eta_k}
{\varepsilon_k}\, ,\hspace{0.3cm}
\varepsilon_k=\sqrt{1-\eta_k^2}\,,\nonumber \\
\eta_k&=&\frac{2\sqrt{\sigma}}{\sigma+1}
\frac{J_{\perp}/2+\gamma_k}{J_{\perp}/2+1}\,,
\hspace{0.3cm} \gamma_k=\cos(ka_0).\nonumber
\end{eqnarray}
The sums run over the wave vectors $k$ from
the lattice Brillouin zone $-\pi/a_0\leq k \leq \pi/a_0$.

${\cal H}_0$ is the usual quadratic quasiparticle  Hamiltonian 
of the linear spin-wave theory (LSWT) {\it corrected}  by
the  diagonal quadratic terms coming from
a normal ordering of
the quartic Dyson-Maleev bosonic interaction.
It is important that these corrections
(similar to Oguchi's corrections in
antiferromagnets\cite{oguchi})
renormalize the magnon excitation spectrum
(and the ground-state energy)
without changing  its  basic  structure,
i.e., the number of Goldstone modes.
In terms of quasiparticle operators
the Hamiltonian ${\cal H}_0$ simply reads
\begin{equation}\label{h0}
{\cal H}_0=2S\sum_k
\omega_k^{(\alpha)}\alpha_k^{\dag}\alpha_k
+\omega_k^{(\beta)}\beta_k^{\dag}\beta_k \, ,
\end{equation}
where
\begin{eqnarray}\label{ek}
\omega_k^{(\alpha,\beta)}&=&\left(1+\frac{J_{\perp}}{2}\right)
\left(\frac{\sigma+1}{2}\varepsilon_k
\mp \frac{\sigma-1}{2}\right) \nonumber \\
&+&\frac{C_k\eta_k-D}{4S\varepsilon_k}
\pm \frac{\sigma-1}{4S\sqrt{\sigma}}
\left( 2c_2+c_3J_{\perp}\right)\,,
\end{eqnarray}
$C_k=c_1(\sigma+1)(2\gamma_k+J_{\perp})/\sqrt{\sigma}+
2(2c_2\gamma_k+c_3J_{\perp})$, and $D=2c_1(2+J_{\perp})
+(\sigma+1)(2c_2+J_{\perp}c_3)/\sqrt{\sigma}$. Here
$\omega_k^{(\alpha,\beta)}$ are the 
{\em dressed} quasiparticle dispersions. The functions
$\omega_k^{(\alpha,\beta)}$ without ${\cal O}(1/S)$ corrections   
will be referred to  as  {\em bare} dispersions. 

Finally, the  quasiparticle interaction $V$
includes two different terms: the  quadratic bosonic
interaction
\begin{equation}
V_2=\sum_kV_k^{(+)}\alpha_k^{\dag}\beta_k^{\dag}
+V_k^{(-)}\alpha_k\beta_k \,,
\end{equation}
which is defined by the vertex functions
\begin{equation}
V_k^{(\pm )}=\frac{D\eta_k-C_k}{2\varepsilon_k}
\mp\frac{\sigma -1}{\sqrt{\sigma}}c_1
\left( \gamma_k+\frac{J_{\perp}}{2}\right)\, ,
\end{equation}
and the quartic  Dyson-Maleev interaction
$V_{DM}$ defined by the vertex functions
$V^{(i)}=V^{(i)}_{12;34}$, $i=1,\ldots ,9$.\cite{note2}
Here and in what follows we use  the wave-vector
abbreviations $(k_1,k_2,k_3,k_4)\equiv (1,2,3,4)$.
\subsection{Perturbation series for $M_A$}    
In spite of the fact that 
the expression for the ground-state energy (\ref{e0}) produces
an excellent fit to the exact-diagonalization (ED) results
in a large interval  up to  $J_{\perp}=10$ 
(in Fig.~\ref{e01} we show a smaller interval),  this fact by itself
is not  enough  to make a statement about the  accuracy of  the SWT. 
It is more appropriate to use as a criterion the sublattice
magnetization $M_A$,  since the latter  keeps  
information on the long-range spin correlations as well.  
The zeroth-order ${\cal O}(V^0)$  result for $m_A=M_A/N$  
simply reads  $m_A=S_1-c_1$. We  shall deal below only with the  
on-site magnetization  $m_A$, as the relation  $m_A+m_B=S_1-S_2$
($m_B=M_B/N$) is fulfilled up to  an arbitrary order of 
the  perturbation series.
Note that in the antiferromagnetic case $S_1=S_2$  
the  off-diagonal interaction
$V_2$ in ${\cal H}_B$ disappears and,
as a result, there are no ${\cal O}(V)$
corrections to $m_A$. In the ferrimagnetic case  $V_2\neq 0$,  
so that the expansion for $m_A$ in powers of $V$ has the form   
\begin{equation}\label{m1}
m_A=S_1-c_1
-\frac{1}{4SN}
\sum_k \frac{\eta_k}{\varepsilon_k}
\frac{V_k^{(+)}+V_k^{(-)}}{\omega_k^{(\alpha)}+\omega_k^{(\beta)}}
+{\cal O}\left( V^2\right)
\,.
\end{equation}

\begin{figure}
\centering\epsfig{file=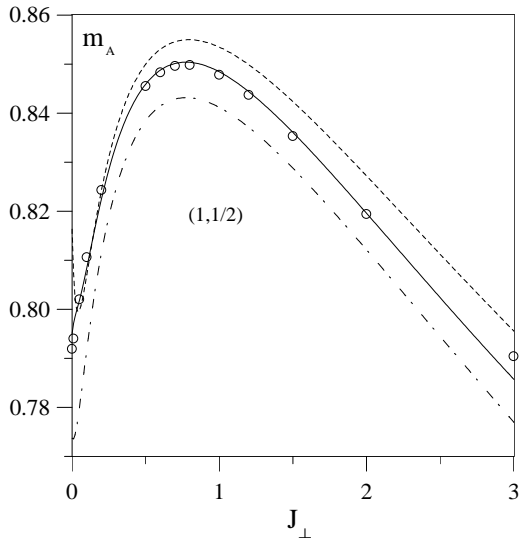,width=7cm}
\vspace{0.3cm}
\narrowtext
\caption{
On-site  magnetization (sublattice ${\cal A}$) of the
ferrimagnetic phase as a function of the interchain coupling
$J_{\perp}$.
The dashed and dashed-dotted  lines  display the
series  results up to first order in powers of $1/S$ (bare
dispersions)  and $V$ (dressed dispersions), respectively 
[Eq.\ (\ref{m1})].
The solid line shows the second-order  ${\cal O}(V^2)$ series
result for $m_A$  (see the Appendix). The Lanczos numerical
results for the $N=12$ ladder  are denoted   by
open circles.
}
\label{m11}
\end{figure}

In Fig.~\ref{m11} we show results for  
$m_A$ obtained from 
Eq.\ (\ref{m1}) by using subsequently the bare and dressed
dispersions $\omega_k^{(\alpha ,\beta)}$ from  Eq.\ (\ref{ek}).
Comparing with the  numerical  results, 
one   indicates that the  expansion in powers of $1/S$ 
(using bare dispersions) does not  describe  qualitatively
the weak-coupling limit $J_{\perp}\ll 1$:
it predicts a small decrease of $m_A$ in the vicinity 
of $J_{\perp}=0$. 
On the other hand, we find  that the expansion in 
powers of $V$ (using dressed  dispersions) 
gives a correct qualitative result in the
weak-coupling limit $J_{\perp}\ll 1$ as well. 
The indicated problem  of the standard $1/S$ expansion
might  be a result of  the enhanced  fluctuations
of  the  chain  magnetizations  
about the common quantization axis,
and/or of   the usual zero-point fluctuations 
[as far as the extreme quantum system $(1,\frac{1}{2})$ 
is concerned]. The above  observation implies that at least in the
extreme quantum case  $(1,\frac{1}{2})$ the expansion
in powers of $V$ is  more reliable: such a viewpoint 
is strongly supported by the second-order ${\cal O}(V^2)$ 
result for $m_A$ (see the Appendix and Fig.~\ref{m11}).   
The above interpretation of the spin-wave series 
(as power series in $V$) is also adopted in  
the following analysis of the   excitation spectrum.  
As a matter of fact, the difference between the standard
$1/S$ expansion  and those in powers of $V$ is mostly
pronounced in the extreme quantum case $(1,\frac{1}{2})$:
for larger site spins  both expansions
practically coincide.  

In the whole interval $0\leq J_{\perp}\leq 3$ shown in 
Fig.~\ref{m11}, the  second-order  result for $m_A$ 
differs by less than $0.3\%$ from the  ED estimates
based on the $N=12$ ladder\cite{note3}. For 
larger $J_{\perp}$  one clearly indicates a monotonic
increase of the deviations from the numerical estimates, which
can be ascribed to the enlarged reduction of the 
classical spin $\delta S_1 =S_1-m_A$: in the limit 
$J_{\perp}\rightarrow \infty$ the exact result is 
$\delta S_1=\frac{1}{3}$.
It is shown in the next Section that   
the SWT description of the excitation spectrum  exhibits 
similar  precision but in a smaller parameter range
 ($0\leq J_{\perp}\leq 2$)  
\section{Excitation spectrum}
\subsection{Free quasiparticles}
\begin{figure}
\centering\epsfig{file=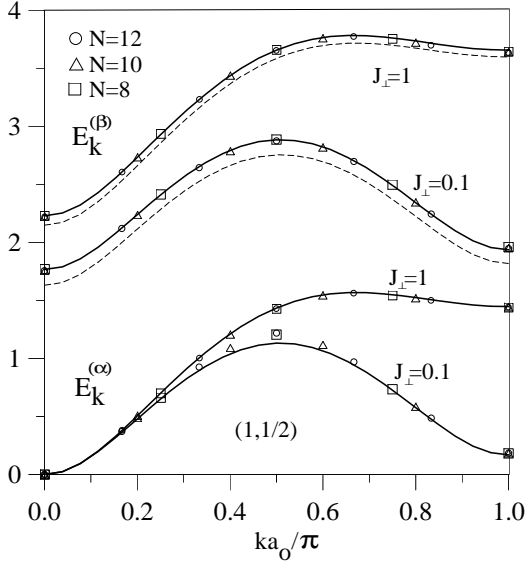,width=7cm}
\vspace{0.3cm}
\narrowtext
\caption{
Excitation spectrum of the
$(1,\frac{1}{2})$ ladder for interchain couplings 
$J_{\perp}=0.1$  and $J_{\perp}=1$.
The dashed and solid  lines display, respectively,  
the series  results up to the orders 
${\cal O}(V^0)\equiv {\cal O}(1/S)$
and  ${\cal O}(V^2)$. 
The ${\cal O}(V^0)$ results for the acoustic  branch
$E_k^{(\alpha )}$ (which are not displayed) closely follow 
the presented second-order curves. 
The symbols indicate our  Lanczos numerical results.
}
\label{eab}
\end{figure}

The quadratic Hamiltonian ${\cal H}_0$
defines two branches of
spin-wave excitations ($\alpha$ and $\beta$ magnons)
described by the dispersion relations 
$E_k^{(\alpha,\beta)}=2S\omega_k^{(\alpha,\beta)}$, 
Eq.\ (\ref{ek}), in  the extended Brillouin zone 
$-\pi/a_0 \leq k \leq \pi/a_0$ (see Fig.~\ref{eab}).
The excited states composed of $\alpha$ and $\beta$ magnons
belong to subspaces with $M^z\leq (S_1-S_2)N-1$ 
and $M^z\geq (S_1-S_2)N+1$, respectively.
The magnon energies $E_k^{(\alpha,\beta)}$ 
are calculated up to the order ${\cal O}(1/S)$.
In the long wavelength limit $ka_0\ll 1$, the ferromagnetic
branch  $E_k^{(\alpha)}$  has the  quadratic Landau-Lifshits form
\begin{equation}\label{landau}
E_k^{(\alpha)}=
\frac{\rho_s}{M_0}k^2+{\cal O}(k^4),
\end{equation}
where  $M_0=(S_1-S_2)/(2a_0)$ and $\rho_s$ are,
respectively, the magnetization
density per chain and the spin-stiffness constant\cite{halperin1}
 of the ferrimagnetic ladder.
This form of the Goldstone modes is characteristic for
Heisenberg ferromagnets and reflects the fact that the
order parameter (i.e., the ferromagnetic moment)
is itself a constant of the motion.\cite{halperin1}
An alternative approach,  relying on the  expected conformal
invariance of the  related XXZ ladder model, may  also be used to
predict the quadratic form of Eq.\ (\ref{landau}).\cite{alcaraz}

The spin-stiffness constant $\rho_s$ and the 
magnetization density $M_0$ play a basic role
in the low-temperature thermodynamics
of 1D models with a continuous  symmetry and  conserved 
order parameters.\cite{read}
Up to the order  ${\cal O}(1/S)$, $\rho_s$ can be obtained
from the Landau-Lifshitz relation (\ref{landau}) and
Eq.\ (\ref{ek}):
\begin{equation}\label{rho0}
\frac{\rho_s}{a_0S_1S_2}=1-c_1\frac{\sigma +1}
{S\sigma}-\frac{c_2}{S\sqrt{\sigma}}\, .
\end{equation}

At the zone boundary $k=\pi/a_0$  the ferromagnetic branch
exhibits an additional  minimum so that in the
vicinity of $\pi/a_0$ the spectrum reads
\begin{equation}\label{ll}
E_k^{(\alpha)}=\Delta_{\pi}^{(\alpha)}+{\rm const}
\left( \frac{\pi}{a_0}-k\right)^2,
\end{equation}
where $\Delta_{\pi}^{(\alpha)}$ is the excitation
gap at the zone boundary.
Such a branch folding of the excitation spectrum is 
typical for ladder structures - it has been indicated in 
uniform spin-$\frac{1}{2}$ ladders as well.\cite{barnes}
The excitation  mode at $k=\pi/a_0$ reflects the underlying
ladder structure and  in the weak-coupling limit $J_{\perp}\ll 1$
may be interpreted as uniform rotations of
the antiferromagnetic moment ${\bf L}=\sum_{n=1}^N (-1)^{n+1}
\left[ {\bf S}_1(n)-{\bf S}_2(n)\right]$.
As  $J_{\perp}\rightarrow 0$, the excitation gap 
$\Delta_{\pi}^{(\alpha)}\propto J_{\perp}$ goes to zero.  
For ferromagnetic couplings $J_{\perp}<0$,
the  $k=\pi/a_0$  mode becomes unstable, thus
producing  a global instability  of the ferrimagnetic state. 
The minimum at $k=\pi/a_0$  persists up to the limit 
$J_{\perp}=\infty$: the related maximum  of
$E_k^{(\alpha)}$ changes its position from $k=\pi/(2a_0)$
(for $J_{\perp}=0$) to $k=\pi/a_0$ (for $J_{\perp}=\infty$).

Similar folding appears in the optical
antiferromagnetic branch $E_k^{(\beta)}$, which can
be characterized by  the spectral gaps $\Delta_0^{(\beta)}$ and
$\Delta_{\pi}^{(\beta)}$ at $k=0$ and $k=\pi$, respectively.
For instance, using Eq.\ (\ref{ek}),  the spectral
gap $\Delta_{0}^{(\beta)}$ up to  ${\cal O}(1/S)$ reads
\begin{equation}\label{delta0}
\Delta_0^{(\beta)}=(2+J_{\perp})(S_1-S_2)\left(1-
\frac{2c_2+c3J_{\perp}}
{S\sqrt{\sigma}(2+J_{\perp})}\right)\,.
\end{equation}
For the $(1,\frac{1}{2})$ ladder  at $J_{\perp}=0$,
Eqs. (\ref{rho0}) and (\ref{delta0})
give the estimates $\rho_s/(a_0S_1S_2)=0.761$ 
and $\Delta_0^{(\beta )}=1.676$.
On the other hand, the LSWT Hamiltonian ${\cal H}_0^{'}$
produces the parameters of the related classical system:
$\rho_s/(a_0S_1S_2)=1$ and $\Delta_0^{(\beta )}=1$.
The  numerical estimate for the gap $\Delta_0^{(\beta )}
=1.759$ at  $J_{\perp}=0$\cite{yamamoto2} clearly demonstrates
the importance of the $1/S$ corrections in Eq.\ (\ref{ek}).
\subsection{Quasiparticle interactions: second-order
corrections for $\omega_k^{(\alpha,\beta)}$}
The quasiparticle interaction $V$ produces corrections to  
the dispersions $\omega_k^{(\alpha,\beta )}$
only up from the order  ${\cal O}(V^2)$.
The second-order corrections for  the ferromagnetic branch
$\omega^{(\alpha)}_k$ are described by the self-energy
diagrams shown in Fig.~2 of Ref.~\onlinecite{ivanov3}.
The respective analytic expression for the corrections reads
\begin{eqnarray}\label{oma2}
&\delta \omega^{(\alpha)}_k&
=-\frac{1}{(2S)^2}\left[ \frac{V^{(+)}_{k}V^{(-)}_{k}}
{\omega^{(\alpha )}_k+\omega^{(\beta )}_k }\right.\nonumber \\
&-&\frac{2}{N}
\sum_{p}\frac{V^{(+)}_{p}V^{(2)}_{kp;pk}+V^{(-)}_{p}V^{(3)}_{kp;pk}}
{\omega^{(\alpha )}_p+\omega^{(\beta )}_p }\nonumber \\
&+&\frac{2}{N^2}
\sum_{2-4}\delta_{k2}^{34} \frac{V^{(8)}_{43;2k}V^{(7)}_{k2;34}}
{\omega^{(\alpha )}_k+\omega^{(\alpha )}_2+\omega^{(\beta )}_3+
\omega^{(\beta )}_4 }\nonumber \\
&+&\left.\frac{2}{N^2}
\sum_{2-4}\delta_{k2}^{34}
\frac{V^{(3)}_{43;2k}V^{(2)}_{k2;34}}
{-\omega^{(\alpha )}_k+\omega^{(\beta )}_2+\omega^{(\alpha )}_3+
\omega^{(\alpha )}_4 }\right]\, .
\end{eqnarray}
Note that since the vertex functions $V_k^{(-)}$, $V^{(2)}_{kp;pk}$,
$V^{(3)}_{kp;pk}$, $V^{(8)}_{43;2k}$, and $V^{(3)}_{43;2k}$
vanish at the zone center $k=0$,
the gapless structure of $\omega_k^{(\alpha)}$ is
preserved separately by each diagram.\cite{note34} $\delta_{12}^{34}
\equiv \delta (1+2-3-4)$ is the Kronecker $\delta$ function.

The second-order corrections
for  $\omega^{(\beta)}_k$ come from similar diagrams.
The explicit expression reads
\begin{eqnarray}\label{omb2}
&\delta \omega^{(\beta)}_k&
= -\frac{1}{(2S)^2}\left[ \frac{V^{(+)}_{k}V^{(-)}_{k}}
{\omega^{(\alpha )}_k+\omega^{(\beta )}_k }\right.\nonumber \\
&-&\frac{2}{N}
\sum_{p}\frac{V^{(+)}_{p}V^{(5)}_{kp;pk}
+V^{(-)}_{p}V^{(6)}_{kp;pk}}
{\omega^{(\alpha )}_p+\omega^{(\beta )}_p }\nonumber \\
&+&\frac{2}{N^2}
\sum_{2-4}\delta_{k2}^{34}\frac{V^{(7)}_{43;2k}
V^{(8)}_{k2;34}}
{\omega^{(\beta )}_k+\omega^{(\beta )}_2+\omega^{(\alpha )}_3+
\omega^{(\alpha )}_4 }\nonumber \\
&+&\left.\frac{2}{N^2}
\sum_{2-4}\delta_{k2}^{34}
\frac{V^{(5)}_{43;2k}V^{(6)}_{k2;34}}
{-\omega^{(\beta )}_k+\omega^{(\alpha )}_2
+\omega^{(\beta )}_3+
\omega^{(\beta )}_4 }\right] \, .
\end{eqnarray}

\begin{figure}
\centering\epsfig{file=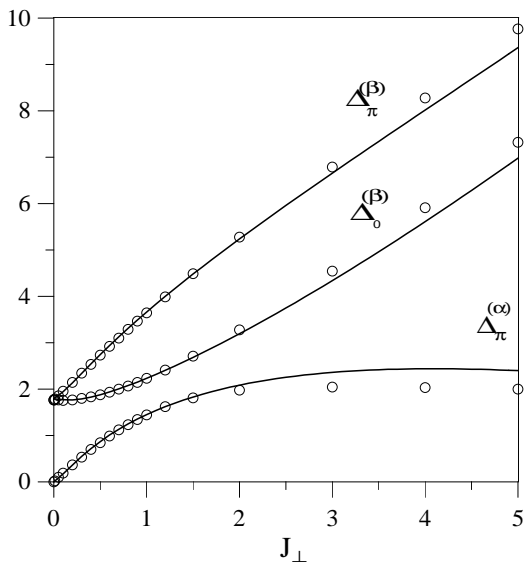,width=7cm}
\vspace{0.3cm}
\narrowtext
\caption{
Gaps in the excitation spectrum of the 
$(1,\frac{1}{2})$ ferrimagnetic  ladder as functions 
of the interchain  coupling $J_{\perp}$. 
The solid lines represent  perturbation  series
results up to second order in $V$.
The open circles indicate
Lanczos numerical results for a ladder with
$N=12$ rungs.
}
\label{gap1}
\end{figure}

The above expressions  can  be used to 
find the second-order corrections 
for the spin-stiffness  constant $\rho_s$ and the excitation 
gaps  $\Delta_{\pi}^{(\alpha)}$, $\Delta_{0}^{(\beta)}$, 
and $\Delta_{\pi}^{(\beta)}$.
The results for the ferrimagnetic state are summarized
in Figs.~\ref{eab}-\ref{rs1} and in Table \ref{table1}.
\section{Analysis of the results and conclusions}
A comparison of the SWT and ED  results (see Fig.~\ref{eab})
implies  that already at the level of free quasiparticles
the theoretical curves for the acoustic branch 
$E_k^{(\alpha)}$ closely follow
the numerical estimates  almost in the whole 
Brillouin zone apart from some  vicinity of the 
wave vector $k=\pi/(2a_0)$.  Note that the indicated  
discrepancy exists  only in  the weak-coupling limit 
(the case $J_{\perp}=0.1$ in  Fig.~\ref{eab}). 
Similar problem has been indicated for  the
$(1,\frac{1}{2})$ ferrimagnetic chain.\cite{ivanov3} 
The point is that  the second-order  corrections 
for $E_k^{(\alpha)}$, Eq.\ (\ref{oma2}),  in the 
vicinity of $k=\pi/(2a_0)$ are very small (about $0.6\%$ 
from the principal approximation 
for $J_{\perp}=0.1$).  
Therefore, in the weak-coupling limit $J_{\perp}\ll 1$ 
the series does not  describe  in a proper way the 
acoustic branch close to the wave vector  $k=\pi/(2a_0)$ 
(where the deviation is about $10\%$ from the extrapolated  
ED  result).
Looking at the recently published Monte Carlo
results for different $(S_1,S_2)$ ferrimagnetic
chains\cite{yamamoto1}, it is seen  that the
discussed discrepancy in the region of  $k=\pi/(2a_0)$ 
is well  pronounced only in   the extreme quantum 
system $(1,\frac{1}{2})$.
On the other hand, turning to the optical branch  
$E_k^{(\beta)}$, one  finds an excellent 
agreement between the second-order  ${\cal O}(V^2)$ 
perturbation result and the numerical estimates
(see Fig.~\ref{eab}).  For instance,  the
theoretical result for the excitation gap $\Delta_0^{(\beta )}$
at $J_{\perp}=0.1$ differs by less than $0.5\%$ from 
the ED estimate. Moreover, it can be expected\cite{ivanov3}
that the third-order corrections further 
improve the above result.\cite{note4}  
The plots of the excitation gaps  
$\Delta_{\pi}^{(\alpha)}$,  
$\Delta_{0}^{(\beta)}$, and $\Delta_{\pi}^{(\beta)}$
(see  Fig.~\ref{gap1}) can  be used to find 
the range where the SWT still produces good
quantitative results. 
The largest discrepancies  are connected
with  the excitation gap $\Delta_{\pi}^{(\alpha)}$ (at
$J_{\perp}=1.5$ the deviation from the ED result is $1.4\%$, 
but at $J_{\perp}=3$ it  already exceeds  $15\%$).
The discrepancies  grow with the  
interchain interaction, 
but up to $J_{\perp}\approx 2$ the theory produces
good quantitative results. 
\begin{figure}
\centering\epsfig{file=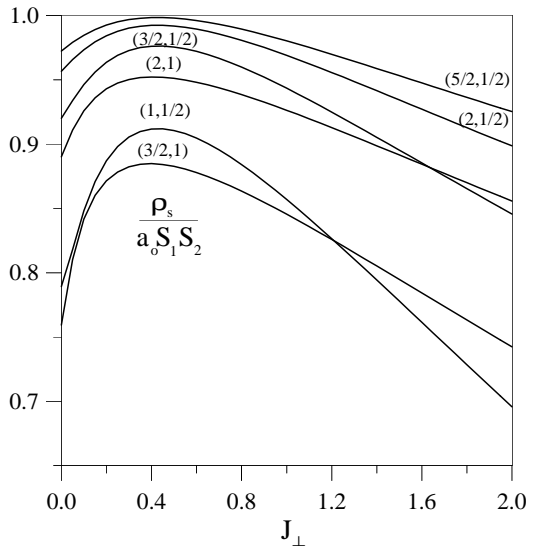,width=7cm}
\vspace{0.3cm}
\narrowtext
\caption{
Spin-stiffness constants versus $J_{\perp}$
of different $(S_1,S_2)$ ferrimagnetic ladders 
calculated up to second order of the perturbation
series in  $V$.
}
\label{rs1}
\end{figure}

The plots of  $m_A$, 
$\Delta_0^{(\beta )}$, and $\rho_s$ (see Figs.~\ref{m11}, 
\ref{gap1}, and \ref{rs1}) also
show   that the short-range correlations,  which are typical for the 
strong-coupling limit $J_{\perp}\rightarrow +\infty$,
are qualitatively established already at $J_{\perp}=2$.  
On the other hand, the same plots indicate    a well 
pronounced crossover behavior for smaller interchain couplings
($0<J_{\perp}<1$): the spin-stiffness constant
$\rho_s$ and  the on-site magnetization $m_A$ reach their 
maximal values (respectively, at $J_{\perp}=0.45$ and $0.77$), 
whereas the excitation gap $\Delta_0^{(\beta )}$ becomes minimal 
(at  $J_{\perp}\approx 0.15$). The discussed  peculiarities 
appear as a result of the crossover
between different types of short-range correlations without
global change  of the  ground-state symmetry. 
       
\begin{figure}
\centering\epsfig{file=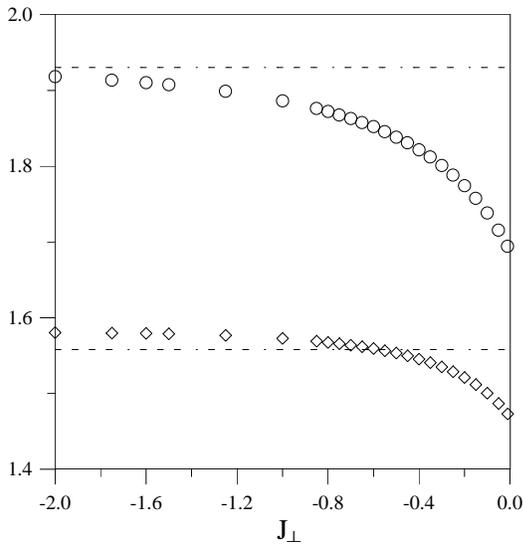,width=7cm}
\vspace{0.3cm}
\narrowtext
\caption{
Short-range intrachain  correlations in the spin-singlet phase
of  the $(1,\frac{1}{2})$  checkerboard ladder.
The diamonds  and circles represent, respectively,
Lanczos numerical results ($N=10$ ladder) for the spin-spin 
correlators $\langle \left[ {\bf S}_1(n)+{\bf S}_2(n)\right]
\cdot \left[ {\bf S}_1(n+2)+{\bf S}_2(n+2)\right]\rangle$
and  
 $\langle \left[ {\bf S}_1(n)+{\bf S}_2(n)\right]
\cdot \left[ {\bf S}_1(n+4)+{\bf S}_2(n+4)\right]\rangle$ 
The dashed lines indicate  ED results ($N=10$)
for the respective spin-spin correlators  in the 
antiferromagnetic spin-$\frac{3}{2}$  
Heisenberg  chain.                           
}
\label{co2}
\end{figure}

\begin{figure}
\centering\epsfig{file=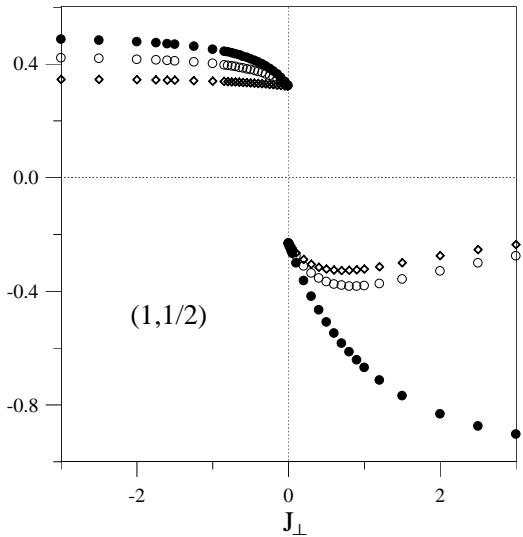,width=7cm}
\vspace{0.3cm}
\narrowtext
\caption{
Short-range interchain  correlations in the 
$(1,\frac{1}{2})$ system, as obtained  from
the numerical diagonalization of the $N=10$ ladder: 
$\langle {\bf S}_1(n)\cdot {\bf S}_2(n)\rangle$-filled circles,
$\langle {\bf S}_1(n)\cdot {\bf S}_2(n+2)\rangle$-open circles,
$\langle {\bf S}_1(n)\cdot {\bf S}_2(n+4)\rangle$-diamonds.                             
}
\label{co1}
\end{figure}

Turning to the case of ferromagnetic 
interchain couplings ($J_{\perp}<0$),  
as mentioned above, one can  expect  
that the model (\ref{h}) is in a disordered  
spin-singlet phase.
In the limit $J_{\perp}\rightarrow -\infty$,  the
system  is equivalent to the antiferromagnetic 
spin-($S_1+S_2$) Heisenberg  chain.
Again the weak-coupling limit  $|J_{\perp}|\ll 1$ is   more
interesting, especially in the presence of exchange 
anisotropies. Here we shall restrict ourselves to short 
comments on the weak-coupling limit
$|J_{\perp}|\ll 1$ of the isotropic $(1,\frac{1}{2})$ 
ladder. The ED results is Figs.~\ref{co2} and \ref{co1} imply 
that already in the weak-coupling limit $|J_{\perp}|\ll 1$ 
the short-range spin correlations qualitatively reproduce
the corresponding  spin correlations in the 
spin-$\frac{3}{2}$ antiferromagnetic Heisenberg  chain.
A qualitative LSWT analysis of the excitation spectrum 
predicts  two degenerate branches of spin-wave 
excitations $E_k^{(\pm)}$. 
The lower branch $E_k^{(-)}$   has a linear dispersion close to the
wave vector $k=0$,  which is characterized by the
spin-wave velocity
\begin{equation}
c_{sw}=\frac{4S_1S_2a_0|J_{\perp}|^{1/2}}
{\sqrt{|J_{\perp}|(S_1+S_2)^2+2(S_1-S_2)^2}}\, .
\end{equation}
On the other hand, $E_k^{(+)}$ describes gapped spin-wave excitations:
the characteristic spectral gap at  $k=0$ is
\begin{equation} 
\Delta = \sqrt{|J_{\perp}|^2(S_1+S_2)^2+|J_{\perp}|(S_1^2+S_2^2)
+4(S_1-S_2)^2}\, . 
\end{equation} 
It is seen that the relativistic modes remain
stable (and the gap $\Delta$
is finite) up to the limit  of noninteracting chains.
The above LSWT result may be interpreted 
in the sense that the classical  magnetic state
in the region $J_{\perp}<0$ is swept out by the quantum 
fluctuations.\cite{langari3} 
On the other hand, as is well known, 
a LSWT analysis  cannot
exclude  the existence of gapped  Haldane-type  phases.
To  some  extend, the discussed 
ferrimagnetic model ($J_{\perp}<0$) in the weak-coupling
limit resembles the  recently studied two-leg spin  
ladder constructed of ferromagnetic  
spin-$\frac{1}{2}$  chains with antiferromagnetic 
interchain couplings 
($J_{\perp}>0$).\cite{kolezhuk,roji}
It was found  that for arbitrary 
$J_{\perp}>0$ and isotropic exchange interactions
the ground state is disordered and the system behaves like
the spin-$1$ Heisenberg antiferromagnetic chain.
Although the above analysis   cannot completely 
exclude the existence of  a similar  gapped  
phase for $|J_{\perp}|\ll 1$, we  believe that the critical phase
remains stable up to  $J_{\perp}=0^-$.
Such a conclusion is supported  
by the ED results which demonstrate  well-established short-range 
correlations even in the limit  $|J_{\perp}|\ll 1$. 
In terms of the related  nonlinear $\sigma$ model description,
on may expect   that one and the same    phase 
$\theta =\pi$ (mod $2\pi$) of the topological term will describe  
the whole parameter range $J_{\perp}<0$. 
\end{multicols}
\widetext    
\begin{table}
\caption{ Results for the spin-stiffness
constant $\rho_s$, the on-site magnetization $m_A$, and
the excitation gaps ($\Delta_{\pi}^{(\alpha)}$, 
$\Delta_{0}^{(\beta)}$, and $\Delta_{\pi}^{(\beta)}$) of  
different $J_{\perp}=1$ ferrimagnetic ladders, as 
obtained from the  second-order 
perturbation series in $V$ .  
}
\label{table1}
\begin{tabular}{ccccccc}
$(S_1,S_2)$&$\rho_s/(a_0S_1S_2)$&$m_A$&$\Delta_{\pi}^{(\alpha)}$&
$\Delta_{0}^{(\beta)}$&$\Delta_{\pi}^{(\beta)}$\\
\tableline
$(1,\frac{1}{2})$&0.8570&0.8484&1.4480&2.2303&3.6596\\
$(\frac{3}{2},1)$&0.8455&1.2517&2.9144&2.0687&4.9666\\
$(\frac{3}{2},\frac{1}{2})$&0.9435&1.3992&1.3987&3.8191&5.2095\\
$(2,1)$&0.9264&1.8285&2.8350&3.6952&6.5197\\ 
$(2,\frac{1}{2})$&0.9691&1.9236&1.3789&5.3623&6.7367\\
$(\frac{5}{2},\frac{1}{2})$&0.9804&2.4382&1.3683&6.8885&8.2540\\  
\end{tabular}
 \end{table}
\begin{multicols}{2}    

In   conclusion, the  accomplished analysis clealy
shows two basic implications of the
underlying lattice structure for the ladder 
ferrimagnetic state: 
(i) the spin-wave  excitations
form folded acoustic and optical branches in the
extended Brillouin zone  and
(ii) the ground state parameters (such as the
on-site magnetizations $m_A$  and the
spin-stiffness constant $\rho_s$) show a 
crossover behavior in the weak-coupling region $0<J_{\perp}<1$.   
The presented results can be used in studies concerning
the low-temperature thermodynamics of ferrimagnetic ladders.

\acknowledgments
We are grateful to J\"org Schulenburg for kindly providing us 
with his  code for numerical diagonalization.   
This work was supported by the Deutsche 
Forschungsgemeinschaft
(Grant No. 436BUL 113/106/1 and  106/2-1) and the 
Bulgarian Science Foundation (Grant No. F817/98).

\appendix
\section{Second-order corrections for $m_A$}
\begin{figure}
\centering\epsfig{file=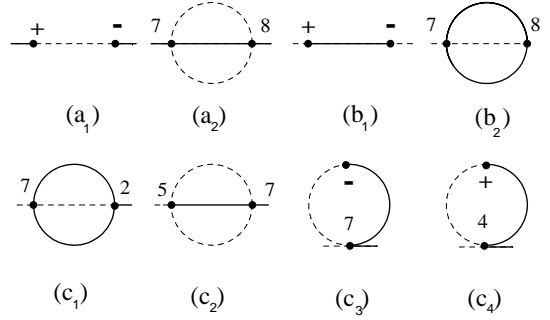,width=7cm}
\vspace{0.3cm}
\narrowtext
\caption{
Second-order diagrams contributing to $m_A^{(2)}$:
(a), (b), and (c) diagrams give contributions,
respectively, 
to $\langle \alpha_k^{\dagger}\alpha_k\rangle^{(2)}$,
$\langle \beta_k^{\dagger}\beta_k\rangle^{(2)}$, and
$\langle \alpha_k\beta_k\rangle^{(2)}$.
The diagrams for $\langle \alpha_k^{\dagger}
\beta_k^{\dagger}\rangle^{(2)}$ can be obtained 
from the (c) diagrams by using the 
following vertex  substitutions: 
$c_1$: $(7,2)\rightarrow (8,3)$, 
$c_2$: $(5,7)\rightarrow (6,8)$,
$c_3$: $(-,7)\rightarrow (+,8)$, and
$c_4$: $(+,4)\rightarrow (-,4)$. 
The  $\alpha$ ($\beta$) magnon propagators
are represented by  solid (dashed) lines, whereas 
the vertex functions are denoted by their
superscript indices.
}
\label{diag1}
\end{figure}

The second-order correction for the on-site 
magnetization
$m_A$ can  be expressed in the form
\begin{equation}\label{ma2}
m_A^{(2)}=-\frac{1}{2N}\sum_k\left<
\alpha_k^{\dagger}\alpha_k
+\beta_k^{\dagger}\beta_k-
\frac{\eta_k}{\varepsilon_k}
\left( \alpha_k^{\dagger}\beta_k^{\dagger}
+\alpha_k\beta_k\right) \right>^{(2)},
\end{equation}
where $\langle A\rangle^{(2)}$ denotes the second-order
correction for  the  ground-state average  of  the operator $A$.
The diagrams  connected to
$\langle \alpha_k^{\dagger}\alpha_k\rangle^{(2)}$,
$\langle\beta_k^{\dagger}\beta_k\rangle^{(2)}$,
$\langle\alpha_k^{\dagger}\beta_k^{\dagger}\rangle^{(2)}$,
and $\langle\alpha_k\beta_k\rangle^{(2)}$ are presented in
Fig.~\ref{diag1}. The related explicit expressions read
\begin{eqnarray}\label{aa}
&&\langle \alpha_k^{\dagger}\alpha_k\rangle^{(2)}
=\frac{1}{(2S)^2}
\frac{V^{(+)}_{k}V^{(-)}_{k}}
{\left(\omega^{(\alpha )}_k
+\omega^{(\beta )}_k \right)^2} \\
&+&\frac{1}{(2S)^2} \frac{2}{N^2}
\sum_{2-4}\delta_{k2}^{34} \frac{V^{(8)}_{43;2k}
V^{(7)}_{k2;34}}{\left( \omega^{(\alpha )}_k+
\omega^{(\alpha )}_2+\omega^{(\beta )}_3
+\omega^{(\beta )}_4\right)^2 },\nonumber
\end{eqnarray}
\begin{eqnarray}\label{bb}
&&\langle \beta_k^{\dagger}\beta_k\rangle^{(2)}
=\frac{1}{(2S)^2}
\frac{V^{(+)}_{k}V^{(-)}_{k}}
{\left(\omega^{(\alpha )}_k
+\omega^{(\beta )}_k \right)^2} \\
&+&\frac{1}{(2S)^2} \frac{2}{N^2}
\sum_{2-4}\delta_{k2}^{34} \frac{V^{(7)}_{43;2k}
V^{(8)}_{k2;34}}{\left( \omega^{(\beta )}_k+
\omega^{(\beta )}_2+\omega^{(\alpha )}_3
+\omega^{(\alpha )}_4\right)^2 }\, ,\nonumber
\end{eqnarray}
\begin{eqnarray}\label{ab}
&&\langle \alpha_k^{\dagger}\beta_k^{\dagger}
+\alpha_k\beta_k\rangle^{(2)}=-\frac{1}{(2S)^2}
\times  \\
&&\left[
\frac{2}{N}
\sum_{q}\frac{V_q^{(-)}V^{(7)}_{kq;qk}+
V_q^{(+)}V^{(8)}_{kq;qk}}
{\left( \omega^{(\alpha )}_k+
\omega^{(\beta )}_k\right)
\left(\omega^{(\alpha )}_k+\omega^{(\alpha )}_q
+\omega^{(\beta )}_k+\omega^{(\beta )}_q\right)}\right. 
\nonumber\\
&+&\frac{2}{N}
\sum_{q}\frac{V_q^{(-)}V^{(7)}_{kq;qk}+
V_q^{(+)}V^{(8)}_{kq;qk}}
{\left( \omega^{(\alpha )}_q+
\omega^{(\beta )}_q\right)
\left(\omega^{(\alpha )}_k+\omega^{(\alpha )}_q
+\omega^{(\beta )}_k+\omega^{(\beta )}_q\right)}\nonumber\\
&+&\frac{2}{N}
\sum_{q}\frac{V_q^{(-)}V^{(4)}_{qk;kq}+
V_q^{(+)}V^{(4)}_{kq;qk}}
{\left( \omega^{(\alpha )}_k+
\omega^{(\beta )}_k\right)
\left(\omega^{(\alpha )}_q
+\omega^{(\beta )}_q\right)}\nonumber\\
&-&\frac{2}{N^2}
\sum_{2-4} \frac{\delta_{k2}^{34}\left( V^{(2)}_{k2;34}
V^{(7)}_{43;2k}+V^{(8)}_{k2;34}
V^{(3)}_{43;2k}\right) }{\left( \omega^{(\alpha )}_k+
\omega^{(\beta )}_k\right)\left( \omega^{(\beta )}_k+
\omega^{(\beta )}_2+\omega^{(\alpha )}_3+
\omega^{(\alpha )}_4 \right) }\nonumber\\
&-&\left. \frac{2}{N^2}
\sum_{2-4} \frac{\delta_{k2}^{34}\left( V^{(7)}_{k2;34}
V^{(5)}_{43;2k}+V^{(6)}_{k2;34}
V^{(8)}_{43;2k}\right) }{\left( \omega^{(\alpha )}_k+
\omega^{(\beta )}_k\right)\left( \omega^{(\alpha )}_k+
\omega^{(\alpha )}_2+\omega^{(\beta )}_3+
\omega^{(\beta )}_4 \right) }\right]\, . \nonumber
\end{eqnarray}


\end{multicols}

\begin{references}
\bibitem{note1} For details concerning the derivation
of the bosonic quasiparticle Hamiltonian, see
Ref.~\onlinecite{ivanov3}.
Note that the magnetic unit cell connected to the
classical ferrimagnetic state contains four spins,
so that in the general case one would need four bosonic
operators. However, due to the model  symmetries,
here it is possible to use only two bosonic operators,
thus considerably reducing the following calculations.
\bibitem{note2} $V_{DM}$ and the related vertex functions
may be obtained from the Dyson-Maleev
interaction of the Heisenber ferrimagnetic chain
(Ref.~\onlinecite{ivanov3}) by using the formal substitution
$\gamma_k\mapsto \gamma_k+J_{\perp}/2$.
\bibitem{note34} The indicated 
gap problem for the acoustic branch $E_k^{(\alpha )}$
in ferrimagnetic chains (Ref.~\onlinecite{yamamoto1})   
is an artifact  of the  non-systematic treatment of 
quasiparticle interactions.
\bibitem{note4} Working  at the level of 
free quasiparticles, one might draw just the 
opposite (incorrect) conclusion  that the spin-wave 
series are more efficient for the ferromagnetic branch.
\end{references}
\end{document}